\documentclass[12pt,tightenlines,nofootinbib,aps,floatfix,pra,%
superscriptaddress,eqsecnum]{revtex4}

\usepackage[utf8]{inputenc}
\usepackage{amsfonts}
\usepackage{amsmath,amssymb}
\usepackage{graphicx}
\usepackage{mathtools}
\usepackage[section]{placeins}
\usepackage{float}
\usepackage[font=footnotesize,labelfont=bf]{caption}
\usepackage{alphalph}
\usepackage{natbib}
\usepackage{textcomp}
\usepackage{amsmath}
\usepackage{amssymb}
\usepackage{multirow}

\providecommand{\U}[1]{\protect\rule{.1in}{.1in}}

\newcommand{\ie}{\text{i.e.}}

\newcommand{\cf}{\text{cf.}}

\makeatletter

\newcommand{\Rmnum}[1]{\expandafter\@slowromancap\romannumeral #1@}
\makeatother

\begin{document}

\title{Causality, universality, and effective field theory \\
for van der Waals interactions}

\author{Serdar Elhatisari}
\email{selhati@ncsu.edu}
\affiliation{Department of Physics, North Carolina State University,
Raleigh, North Carolina 27695, USA}

\author{Sebastian K\"onig}
\email{koenig@hiskp.uni-bonn.de}
\affiliation{Helmholtz-Institut f\"ur Strahlen- und Kernphysik (Theorie) and
Bethe Center for Theoretical Physics, Universit\"at Bonn, 53115 Bonn, Germany}

\author{Dean Lee}
\email{dean_lee@ncsu.edu}
\affiliation{Department of Physics, North Carolina State University,
Raleigh, NC 27695, USA}

\author{H.-W. Hammer}
\email{hammer@hiskp.uni-bonn.de}
\affiliation{Helmholtz-Institut f\"ur Strahlen- und Kernphysik (Theorie) and
Bethe Center for Theoretical Physics, Universit\"at Bonn, 53115 Bonn, Germany}

\date{\today}

\begin{abstract}
We analyze low-energy scattering for arbitrary short-range interactions plus
an attractive $1/r^{6}$ tail.  We derive the constraints of causality and
unitarity and find that the van der Waals length scale dominates over
parameters characterizing the short-distance physics of the interaction.
This separation of scales suggests a separate universality class for physics
characterizing interactions with an attractive $1/r^{6}$ tail.  We argue that
a similar universality class exists for any attractive potential $1/r^{\alpha}$
for $\alpha\geq2$.  We also discuss the extension to multichannel systems near
a magnetic Feshbach resonance.  We discuss the implications for
effective field theory
with attractive singular power-law tails.
\end{abstract}

\maketitle

\section{Introduction}

\label{sec:introduction}

Low-energy universality appears when there is a large separation between the
short-distance scale of the interaction and the physically relevant
long-distance scales.  Some well-known examples include the unitarity limit
of two-component
fermions~\cite{O'Hara:2002,Kinast:2004,Zwierlein:2004,Bartenstein:2004,Ku:2012a}
and the Efimov effect in three-body and four-body
systems~\cite{Efimov70,Efimov73,Bedaque:1998kg,
Platter:2004pra,Hammer:2006ct,vonStecher:2008a,PhysRevA.82.040701,
Hadizadeh:2011qj,PhysRevA.84.052503,Kraemer:2006Nat}.
See Refs.~\cite{Braaten:2004a,Giorgini:2007a} for reviews of the subject and
literature.  There have been many theoretical studies of low-energy phenomena
and universality for interactions with finite range.  These studies have direct
applications to nuclear physics systems such as cold dilute neutron matter or
light nuclei such as the triton and alpha particle.  To a good approximation,
the van der Waals interactions between alkali-metal atoms can also be treated as a
finite-range interaction.

However, there are some differences.  For potentials
with an attractive $1/r^{\alpha}$ tail and $\alpha > 2$, the $s$-wave scattering
phase shift near threshold has been formulated in Ref.~\cite{PhysRevA.84.032701}.
For $\alpha > 3$, the modified scattering parameters for an $s$-wave Feshbach
resonance were derived in Ref.~\cite{PhysRevA.85.052703} using coupled-channel
calculations. Analytical expressions for the $s$-wave scattering length and effective
range for two neutral atoms and $\alpha = 6$ have been derived in Ref.~\cite{PhysRevA.59.1998}. However, the applicability of the effective range theory is limited for interactions with
attractive tails. In order to define the scattering length for angular momentum
$L\geq2$ and the effective range for $L\geq1$, a modified version of effective range theory known as quantum-defect theory is needed~\cite{PhysRevA.58.4222,PhysRevA.80.012702}.
Furthermore, scattering parameters of magnetically tunable multichannel systems
have been studied in the context of multichannel quantum-defect
theory~\cite{PhysRevA.70.012710, PhysRevA.72.042719}. See Ref.~\cite{PhysRevA.87.032706}
for a very recent development of multichannel quantum-defect theory for higher partial waves.
There is also growing empirical evidence that there exists a new type of low-energy universality
that ties together all interactions with an attractive $1/r^{6}$ tail. This
might seem surprising since there is no such analogous behavior for
interactions with a Coulomb tail. In this paper we derive the theoretical
foundations for this van der Waals universality at low energies by studying
the near-threshold behavior and the constraints of causality.  We also show
that this universality extends to any power-law interaction $1/r^{\alpha}$
with $\alpha\geq2$ in any number of dimensions.  Our analysis applies to
energy-independent interactions.  We first consider a single
scattering channel but then also consider multichannel systems near a
magnetic Feshbach
resonance.  The full analysis for the multichannel problem off resonance
will be discussed in future publications.

In our analysis we assume that the two-body potential has a long-distance
attractive tail of the form $-C_{6}/r^{6}$.  We define the van der Waals
length scale, $\beta_{6}$, as
\begin{equation}
 \beta_{6}=(2\mu C_{6})^{\frac{1}{4}} \,,
\label{vanderWaalslength}
\end{equation}
where $\mu$ is the reduced mass of the scattering particles.  For simplicity
we use atomic units (a.u.) throughout our discussion.  So, in particular,
we set $\hbar=1$.  In Refs.~\cite{CalleCordon:2010a,RuizArriola:2011a} it was
noticed that an approximate universal relationship exists between the
effective range and inverse scattering length for $s$-wave scattering in many
different pairs of scattering alkali-metal atoms.  If we write $A_{0}$ as the
scattering length and $R_{0}$ as the effective range, the relation is
\begin{equation}
 R_{0}\approx\frac{\beta_{6}\Gamma(1/4)^{2}}{3\pi}-\frac{4\beta_{6}^{2}}
 {3A_{0}}+\frac{8\pi\beta_{6}^{3}}{3\Gamma(1/4)^{2}A_{0}^{2}} \,.
\label{arriola}
\end{equation}
This approximate relation becomes exact for a pure $-C_{6}/r^{6}$ potential.
What is surprising about Eq.~(\ref{arriola}) is that the van der Waals
length $\beta_{6}$ dominates over other length scales which characterize the
short-distance repulsive force between alkali-metal atoms.  This approximate
universality suggests there is some separation of scales between the van der
Waals length $\beta_{6}$ and the length scales of the short-range forces.
This separation of scales will become more transparent later in our analysis
when we determine the coefficients of the short-range $K$ matrix.  It would
be useful to exploit
the separation of scales as an effective field theory with an explicit
van der Waals tail plus contact interactions.  In this paper, we discuss the
constraints on such a van der Waals effective field theory.

We note that a similar dominance of the van der Waals length $\beta_{6}$ has
been discovered for the three-body parameter in the Efimov
effect~\cite{Berninger:2011a,Wang:2012a,Wang:2012b}.  In this paper we focus
only on two-body systems.  However, our analysis should be useful in developing
the foundations for van der Waals effective field theory.  This in turn could
be used to investigate the Efimov effect and other low-energy phenomena in a
model-independent way. An extension of our analysis may be useful to
understand the recently observed universality of
the three-body parameter for narrow Feshbach resonances
\cite{Roy2013}.

The organization of our paper is as follows.  We first discuss the connection
between causality bounds and effective field theory.  Next we consider
asymptotic solutions of the Schr\"{o}dinger equation.  After that, we derive
causality bounds for the short-range $K$ matrix and consider the impact of
these results on van der Waals effective field theory.  Then we discuss
quantum-defect theory and calculate causal ranges for several examples of
single-channel $s$-wave scattering in alkali-metal atoms.  We also consider the
constraints of causality near magnetic Feshbach resonances.  We then
conclude with a summary and discussion.

\section{Causality bounds and effective field theory}

For an effective field theory with local contact interactions, the range of
the interactions is controlled by the momentum cutoff scale.  Problems with
convergence can occur if the cutoff scale is set higher than the scale of the
new physics not described by the effective theory.  It is useful to have a
quantitative measure of when problems may or may not appear, and this is where
the causality bound provides a useful diagnostic tool.  For each scattering
channel we use the physical scattering parameters to compute a quantity called
the causal range, $R^{b}$, which is the minimum range for the interactions
consistent with the requirements of causality and unitarity.  For any fixed
cutoff scale, the causality bound marks a branch cut of the effective theory
when viewed as a function of physical scattering
parameters~\cite{EurPhysJA.48.110,Koenig:2012bv}.  The coupling constants of
the effective theory become complex for scattering parameters violating the
causality bound.  These branch cuts do not appear in perturbation theory, but
they can spoil the convergence pattern of the perturbative expansion.

Wigner was the first to recognize the constraints of causality and unitarity
for two-body scattering with finite-range interactions~\cite{Wigner:1955a}.
The time delay of a scattered wave packet is given by the energy derivative of
the phase shift,
\begin{equation}
 \Delta t=2\frac{d\delta}{dE} \,.
\end{equation}
It is clear that the incoming wave packet must first reach the interacting
region before the outgoing wave packet can leave.  So the causality bound can
be viewed as a lower bound on the time delay, $\Delta t$.  When applied to
wave packets near threshold, the causality bound becomes an upper bound on
the effective range parameter.  Phillips and Cohen derived this bound for
$s$-wave scattering with finite-range interactions~\cite{Phillips:1996ae}. Some
constraints on nucleon-nucleon scattering and the chiral two-pion exchange
potential were considered in Ref.~\cite{PavonValderrama:2005wv}, and relations
between the scattering length and effective range have been explored for
one-boson exchange potentials~\cite{Cordon:2009pj}.  As mentioned above, the
same authors studied the relationship between the scattering length and
effective range for the van der Waals
interaction~\cite{CalleCordon:2010a,RuizArriola:2011a}

In Refs.~\cite{Hammer:2009zh,Hammer:2010fw} the causality and unitarity bounds
for finite-range interactions were extended to an arbitrary number of space-time
dimensions or value of angular momentum.  A complementary discussion based
upon conformal symmetry and scaling dimensions can be found in
Ref.~\cite{Nishida:2010a}.  Coupled-channel systems with partial-wave mixing
were first studied in Ref.~\cite{EurPhysJA.48.110}, and the interactions with
attractive and repulsive Coulomb tails were first considered in
Ref.~\cite{Koenig:2012bv}.

\section{Asymptotic solutions of the Schr\"{o}dinger equation}

We consider a system of two spinless particles interacting via a spherically
symmetric potential in the center-of-mass frame.  As noted in the
Introduction, we use atomic units where $\hbar=1$.  The total wave function in
the relative coordinate $\vec{r}$ can be separated into the radial and the
angular parts as
\begin{equation}
 \Psi_{L,M}^{(k)}(\vec{r})=R_{L}^{(k)}(r)\,Y_{L}^{M}(\hat{r}) \,,
\end{equation}
where $k$ is the magnitude of the spatial momentum and $Y_{L}^{M}(\hat{r})$
are the spherical harmonics.  We define the rescaled radial wave function
$U_{L}^{(k)}(r)$ as
\begin{equation}
 U_{L}^{(k)}(r)=r\,R_{L}^{(k)}(r) \,.
\label{eqn:rescaled_wave_function_1a}
\end{equation}

We first review the case where the only interaction between the two particles
has a finite range, $R$.  This means that the interactions are exactly zero
when the two particles exceed a distance $R$.  Let $\mu$ denote the reduced
mass of the two-body scattering system.  The radial Schr\"{o}dinger equation
for a scattering state with energy $E=k^{2}/(2\mu)$ is then
\begin{equation}
 \left[\frac{d^{2}}{dr^{2}}-\frac{L(L+1)}{r^{2}}+k^{2}\right] U_{L}^{(k)}(r)
 =2\mu\int_{0}^{R}dr^{\prime}\,W(r,r^{\prime})U_{L}^{(k)}(r^{\prime}) \,.
\end{equation}
We have written the interaction as a rotationally invariant operator with real
kernel $W(r,r^{\prime})$ to avoid any assumption regarding the locality or
nonlocality of the potential.  In our analysis we consider only interactions
which are energy independent.  The finite-range condition implies that
$W(r,r^{\prime})=0$ if $r>R$ or $r^{\prime}>R$.  We assume that the
interaction is sufficiently well behaved at the origin to admit a regular
solution.  This assumption imposes the restriction that at short distances
the potential is not too singular such that the radial wave function satisfies
the regularity condition
\begin{equation}
 \lim_{r\rightarrow0}U_{L}^{(k)}(r)\frac{d}{dr}{U_{L}^{(k)}}(r)=0 \,.
\label{regularity}
\end{equation}
In Ref.~\cite{PhysRevA.30.1279} it is proven that this condition is fulfilled
by a class of potentials $V(r)$ provided that
\begin{equation}
 \int_{0}^{R}r^{\prime}\left|V(r^{\prime})\right| dr^{\prime} < \infty \,.
\end{equation}
We choose a normalization such that, for $r>R$, the radial wave function has the
form
\begin{equation}
 U_{L}^{(k)}(r)=k^{L+\frac{1}{2}}\sqrt{\frac{\pi r}{2}}\left[\cot\delta_{L}(k)
 J_{L+\frac{1}{2}}(kr)-N_{L+\frac{1}{2}}(kr)\right] \,,
\end{equation}
where $J_{L+\frac{1}{2}}(kr)$ and $N_{L+\frac{1}{2}}(kr)$ are the Bessel
functions of the first and second kind, and $\delta_{L}(k)$ is the scattering
phase shift.

We now discuss the main problem of interest where the interactions have a
long-range van der Waals tail.  In addition to the non-singular finite-range
interaction parameterized by $W(r,r^{\prime})$, we assume that there is a
long-range local potential $-C_{6}/r^{6}$ for $r>R$.  The van der Waals
length scale $\beta_{6}$ was defined in Eq.~\eqref{vanderWaalslength}.  The
radial Schr{\"{o}}dinger equation is
\begin{equation}
 \left[\frac{d^{2}}{dr^{2}}-\frac{L(L+1)}{r^{2}}+\frac{\beta_{6}^{4}}{r^{6}}
 \theta(r-R)+k^{2}\right] U_{L}^{(k)}(r)
 = 2\mu\int_{0}^{R}dr^{\prime}\,W(r,r^{\prime})U_{L}^{(k)}(r^{\prime}) \,.
\label{Schrodinger_van_der_Waals}
\end{equation}
The step function $\theta(r-R)$ cuts off the long-range potential at distances
less than $R$.  This ensures that we satisfy the regularity condition in
Eq.~\eqref{regularity} and avoids mathematical problems~associated with
unregulated singular potentials \cite{RevModPhys.43.36}.  The general form of
the solutions for Eq.~\eqref{Schrodinger_van_der_Waals} has been discussed by
Gao in Ref.~\cite{PhysRevA.78.012702}.

In order to simplify some of the more lengthy expressions to follow, we
introduce dimensionless rescaled variables $r_{s}={r}/{\beta_{6}}$,
$k_{s}=\beta_{6}k$, and $\rho_{s}=1/(2r_{s}^{2})$.  In the outer region,
$r>R$, the Schr{\"{o}}dinger equation reduces to
\begin{equation}
 \left[\frac{d^{2}}{dr^{2}}-\frac{L(L+1)}{r^{2}}
 + \frac{\beta_{6}^{4}}{r^{6}}+k^{2}\right] U_{L}^{(k)}(r)=0 \,
\end{equation}
or
\begin{equation}
 \left[\frac{d^{2}}{dr_{s}^{2}}-\frac{L(L+1)}{r_{s}^{2}}+\frac{1}{r_{s}^{6}}
 + k_{s}^{2}\right]  U_{L}^{(k)}(r)=0 \,.
\label{pure_vdW}
\end{equation}
The exact solutions for Eq.~(\ref{pure_vdW}) have been studied in detail in
Ref.~\cite{PhysRevA.58.1728} using the formalism of quantum-defect
theory~\cite{RepProgPhys.46.167,PhysRevA.19.1485,PhysRevA.26.2441}.

The van der Waals wave functions $F_{L}$ and $G_{L}$ are linearly independent
solutions of Eq.~(\ref{pure_vdW}).  In order to write these out we first
need several functions defined in
Appendix~\ref{append:analytic_solution_vdWaals_eq}.  The van der Waals
wave functions $F_{L}$ and $G_{L}$ can be written as summations of Bessel
functions,
\begin{multline}
 {F}_{L}(k,r)=\frac{r_{s}^{1/2}}{X_{L}^{2}(k_{s})+Y_{L}^{2}(k_{s})}\Bigg[
 X_{L}(k_{s})\sum_{m=-\infty}^{\infty}b_{m}(k_{s})\,J_{\nu+m}
 \left(\rho_{s}\right) \\ - Y_{L}(k_{s})\sum_{m=-\infty}^{\infty}b_{m}(k_{s})
 \,N_{\nu+m}\left(\rho_{s}\right)\Bigg] \,,
\label{F_L}
\end{multline}
\begin{multline}
 {G}_{L}(k,r)=\frac{r_{s}^{1/2}}{X_{L}^{2}(k_{s})+Y_{L}^{2}(k_{s})}\Bigg[
 X_{L}(k_{s})\sum_{m=-\infty}^{\infty}b_{m}(k_{s})\,N_{\nu+m}
 \left(\rho_{s}\right) \\ + Y_{L}(k)\sum_{m=-\infty}^{\infty}b_{m}(k_{s})
 \,J_{\nu+m}\left(\rho_{s}\right)\Bigg] \,.
\label{G_L}
\end{multline}
The function $X_{L}$ is defined in Eq.~(\ref{X_L}), and $Y_{L}$ is defined
in Eq.~(\ref{Y_L}).  For $m\geq0$ the function $b_{m}$ is given in
Eq.~(\ref{bm}), while $b_{-m}$ is given in Eq.~(\ref{bmm}).  The offset
$\nu$ appearing in the order of the Bessel functions is given by the solution
of Eq.~(\ref{v}) in Appendix~\ref{append:analytic_solution_vdWaals_eq}.
For notational convenience, however, we omit writing the explicit $k_{s}$
dependence of $\nu$.  Let us define $\delta_{L}^{(\text{short})}(k)$ to be
the phase shift of the van der Waals wave functions due to the scattering from
the short-range interaction.  The normalization of $U_{L}^{(k)}(r)$ is chosen
so that, for $r>R$,
\begin{equation}
 U_{L}^{(k)}(r)=F_{L}(k,r)-\tan\delta_{L}^{(\text{short})}(k)\,G_{L}(k,r) \,.
\label{U_L}
\end{equation}
Our van der Waals wave functions are related to the functions $f_{L}^{c0}$ and
$g_{L}^{c0}$ defined of Ref.~\cite{PhysRevA.78.012702} by the normalization
factors $F_{L}=f_{L}^{c0}/\sqrt{2}$ and $G_{L}=-g_{L}^{c0}/\sqrt{2}$.
Henceforth, we write all expressions in terms of the short-range reaction
matrix
\begin{equation}
 \hat{K}_{L}=\tan\delta_{L}^{(\text{short})}(k) \,,
\end{equation}
which is related to
the short-range scattering matrix via
\begin{equation}
 \hat{S}_{L}=e^{2i\delta_{L}^{(\text{short})}}
 = \frac{i-\hat{K}_{L}}{i+\hat{K}_{L}} \,.
\end{equation}

For any finite-range interaction, $\hat{K}_{L}$ is analytic in $k^{2}$ and can
be calculated by matching solutions for $r\leq R$ and $r>R$ at the boundary.
It can be written in compact form as
\begin{equation}
 \hat{K}_{L}
 =\left.\frac{W(U_{L}^{(k)},F_{L}^{(k)})}{W(U_{L}^{(k)},
 G_{L}^{(k)})}\right\vert _{r=R},
\end{equation}
where $U_{L}^{(k)}$ is the solution of Eq.~(\ref{Schrodinger_van_der_Waals})
that is regular at the origin, and $W$ denotes the Wronskian of two functions,
\begin{equation*}
 W(f,g)=fg^{\prime}-f^{\prime}g \,.
\end{equation*}

\section{Causality bounds for short-range $K$-matrix $\hat{K}_{L}$}

In this section we derive causality bounds for the short-range $K$ matrix
$\hat{K}_{L}$.  For this we need to expand the wave function $U_{L}^{(k)}(r)$
in powers of $k^{2}$.  The steps we follow are analogous to those used in
Refs.~\cite{Hammer:2009zh,Hammer:2010fw,EurPhysJA.48.110,Koenig:2012bv}.  We
first expand $\hat{K}_{L}$,
\begin{equation}
 \hat{K}_{L}=\tan\delta_{L}^{(\text{short})}(k)
 =\sum_{n=0}^{\infty}K_{L,2n}\,k^{2n} \,.
\label{eq:K_L_expansion}
\end{equation}
The first two terms $K_{L,0}$ and $K_{L,2}$ are analogous to the inverse
scattering length and effective range parameters in the usual effective range
expansion.  The higher-order terms can be regarded as analogs of the shape
parameters.  Next we expand the van der Waals wave functions in powers of
$k^{2}$,
\begin{equation}
 {F}_{L}(k,r)=f_{L,0}(r)+f_{L,2}(r)\,k^{2}+{O}(k^{4}) \,,
\label{F_L_expansion}
\end{equation}
\begin{equation}
 {G}_{L}(k,r)=g_{L,0}(r)+g_{L,2}(r)\,k^{2}+{O}(k^{4}) \,.
\label{G_L_expansion}
\end{equation}
In the following, we define
\begin{equation*}
 \nu_{0}=\frac{1}{4}(2L+1),
\end{equation*}
which corresponds to the value of $\nu$ at threshold.  Using the low-energy
expansions in Appendix~\ref{low_energy_expansion}, we find that the
coefficients in Eq.~(\ref{F_L_expansion}) are
\begin{equation}
f_{L,0}(r)=r_{s}^{1/2}J_{\nu_{0}}\left(\rho_{s}\right) \,
\end{equation}
and
\begin{multline}
 f_{L,2}(r)=\frac{\Gamma(\nu_{0})\Gamma(2\nu_{0}-1)}{\Gamma(\nu_{0}+1)
 \Gamma(2\nu_{0})}\frac{\beta_{6}^{2}}{16}r_{s}^{1/2}\big[J_{\nu_{0}-1}
 \left(\rho_{s}\right)+N_{\nu_{0}}\left(\rho_{s}\right)\big] \\
 -\frac{\Gamma(\nu_{0})\Gamma(2\nu_{0}+1)}{\Gamma(\nu_{0}+1)\Gamma(2\nu_{0}+2)}
 \frac{\beta_{6}^{2}}{16}r_{s}^{1/2}\big[J_{\nu_{0}+1}\left(\rho_{s}\right)
 -N_{\nu_{0}}\left(\rho_{s}\right)\big] \,.
\label{eqn:second_term_pair_func_F_1a}
\end{multline}
Similarly, the coefficients in Eq.~(\ref{G_L_expansion}) are
\begin{equation}
 g_{L,0}(r)=r_{s}^{1/2}N_{\nu_{0}}\left(\rho_{s}\right) \,
\end{equation}
and
\begin{multline}
 g_{L,2}(r)=\frac{\Gamma(\nu_{0})\Gamma(2\nu_{0}-1)}{\Gamma(\nu_{0}+1)
 \Gamma(2\nu_{0})}\frac{\beta_{6}^{2}}{16}r_{s}^{1/2}\big[N_{\nu_{0}-1}
 \left(\rho_{s}\right)-J_{\nu_{0}}\left(\rho_{s}\right)\big] \\
 -\frac{\Gamma(\nu_{0})\Gamma(2\nu_{0}+1)}{\Gamma(\nu_{0}+1)\Gamma(2\nu_{0}+2)}
 \frac{\beta_{6}^{2}}{16}r_{s}^{1/2}\big[N_{\nu_{0}+1}\left(\rho_{s}\right)
 +J_{\nu_{0}}\left(\rho_{s}\right)\big] \,.
\label{eqn:second_terms_pair_func_G_1a}
\end{multline}
Using Eq.~(\ref{U_L}), we can now express $U_{L}^{(k)}(r)$ as an expansion in
powers of $k^{2}$.  For $r>R$, we have
\begin{multline}
 U_{L}^{(k)}(r)=f_{L,0}(r)-K_{L,0} g_{L,0}(r) \\
 +k^{2}\big[f_{L,2}(r)-K_{L,0} g_{L,2}(r)-K_{L,2} g_{L,0}(r)\big]
 +{O}(k^{4}) \,.
\label{eqn:wave_function_vdWaals_2a}
\end{multline}

We now consider two solutions of the Schr{\"{o}}dinger equation
$U_{L}^{(k_{a})}(r)$ and $U_{L}^{(k_{b})}(r)$ with momenta $k_{a}$ and $k_{b}$,
respectively.  We have
\begin{equation}
 \left[\frac{d^{2}}{dr^{2}}-\frac{L(L+1)}{r^{2}}+\frac{\beta_{6}^{4}}{r^{6}}
 \theta(r-R)+k_{a}^{2}\right]  U_{L}^{(k_{a})}(r)=2\mu\int_{0}^{R}dr^{\prime}
 \,W(r,r^{\prime})U_{L}^{(k_{a})}(r^{\prime}) \,,
\end{equation}
\begin{equation}
 \left[\frac{d^{2}}{dr^{2}}-\frac{L(L+1)}{r^{2}}+\frac{\beta_{6}^{4}}{r^{6}}
 \theta(r-R)+k_{b}^{2}\right]  U_{L}^{(k_{b})}(r)=2\mu\int_{0}^{R}dr^{\prime}
 \,W(r,r^{\prime})U_{L}^{(k_{b})}(r^{\prime}) \,.
\end{equation}
Following the same steps as in Eq.~(31)--(36) of Ref.~\cite{Hammer:2010fw}, we
obtain the Wronskian integral formula
\begin{equation}
 \frac{W[U_{L}^{(k_{b})},U_{L}^{(k_{a})}](r)}{k_{b}^{2}-k_{a}^{2}}=\int_{0}^{r}
 U_{L}^{(k_{a})}(r^{\prime})U_{L}^{(k_{b})}(r^{\prime})\,dr^{\prime} \,,
\label{eqn:wronskian_integral_relation_1a}
\end{equation}
for any $r>R$.  Using Eq.~(\ref{eqn:wave_function_vdWaals_2a}) for momenta
$k_{a}$ and $k_{b}$ we find
\begin{equation}
\begin{split}
 \frac{W[U_{L}^{(b)},U_{L}^{(a)}](r)}{k_{b}^{2}-k_{a}^{2}}
 &=W[f_{L,2},f_{L,0}](r)-K_{L,0}\big\{
 W[g_{L,2},f_{L,0}](r)+W[f_{L,2},g_{L,0}](r)\big\} \\
 &\hspace{2em}+K_{L,0}^{2}W[g_{L,2},g_{L,0}](r)-K_{L,2}W[g_{L,0},f_{L,0}](r)
 +{O}(k_{a}^{2},k_{b}^{2}) \,.
\end{split}
\label{eqn:wronskian_of_wave_functions_1a}
\end{equation}
In the Wronskian integral formula~\eqref{eqn:wronskian_integral_relation_1a},
we set $k_{a}=0$ and take the limit $k_{b}\rightarrow0$.  With the
wave function at zero energy written as $U_{L}^{(0)}$, the result is
\begin{equation}
 K_{L,2}=b_{L}(r)-\frac{\pi}{4}\int_{0}^{r}
 \left[U_{L}^{(0)}(r^{\prime})\right]^{2}\,dr^{\prime} \,,
\label{eqn:causality_bound_1a}
\end{equation}
where
\begin{equation}
\begin{split}
 b_{L}(r) = &\frac{\pi}{4}W[f_{L,2},f_{L,0}](r)+\frac{\pi}{4}
 K_{L,0}^{2}W[g_{L,2},g_{L,0}](r) \\
 &-\frac{\pi}{4}K_{L,0}
 \big\{W[g_{L,2},f_{L,0}](r)+W[f_{L,2},g_{L,0}](r)\big\} \,.
\end{split}
\label{eqn:causality_bound_b_function_1a}
\end{equation}
The Wronskians appearing in Eq.~(\ref{eqn:causality_bound_b_function_1a}) can
be written out explicitly as
\begin{align}
 W[f_{L,2},f_{L,0}](r) &=\frac{\beta_{6}\rho_{s}}{16\nu_{0}(2\nu_{0}-1)}
 \left[J_{\nu_{0}-2}(\rho_{s})\ J_{\nu_{0}}(\rho_{s})
 -J_{\nu_{0}-1}^{2}(\rho_{s})\right] \nonumber \\
 &+\frac{\beta_{6}\rho_{s}}{16\nu_{0}(2\nu_{0}+1)}\left[J_{\nu_{0}+2}(\rho_{s})
 \ J_{\nu_{0}}(\rho_{s})-J_{\nu_{0}+1}^{2}(\rho_{s})\right] \nonumber \\
 &+\frac{\beta_{6}\rho_{s}}{4(2\nu_{0}-1)(2\nu_{0}+1)}
 \left[J_{\nu_{0}+1}(\rho_{s})
 \ J_{\nu_{0}-1}(\rho_{s})-J_{\nu_{0}}^{2}(\rho_{s})
 +\frac{4}{\pi\rho_{s}}\right] \,,
\label{Wff}
\end{align}
\begin{align}
 W[g_{L,2},g_{L,0}](r) &= \frac{\beta_{6}\rho_{s}}{16\nu_{0}(2\nu_{0}-1)}
 \left[N_{\nu_{0}-2}(\rho_{s})\ N_{\nu_{0}}(\rho_{s})
 -N_{\nu_{0}-1}^{2}(\rho_{s})\right] \nonumber \\
 &+\frac{\beta_{6}\rho_{s}}{16\nu_{0}(2\nu_{0}+1)}\left[N_{\nu_{0}+2}(\rho_{s})
 \ N_{\nu_{0}}(\rho_{s})-N_{\nu_{0}+1}^{2}(\rho_{s})\right] \nonumber \\
 &+\frac{\beta_{6}\rho_{s}}{4(2\nu_{0}-1)(2\nu_{0}+1)}
 \left[N_{\nu_{0}+1}(\rho_{s})\ N_{\nu_{0}-1}(\rho_{s})
 -N_{\nu_{0}}^{2}(\rho_{s})+\frac{4}{\pi\rho_{s}}\right] \,,
\label{Wgg}
\end{align}
and
\begin{multline}
 W[g_{L,2},f_{L,0}](r) = W[f_{L,2},g_{L,0}](r) \\
 =\frac{\beta_{6}\rho_{s}}{16\nu_{0}(2\nu_{0}-1)}\big\{J_{\nu_{0}-1}(\rho_{s})
 \left[N_{\nu_{0}+1}(\rho_{s})-N_{\nu_{0}-1}(\rho_{s})\right]
 -N_{\nu_{0}}(\rho_{s})\left[J_{\nu_{0}}(\rho_{s})
 -J_{\nu_{0}-2}(\rho_{s})\right]\big\} \\
 -\frac{\beta_{6}\rho_{s}}{16\nu_{0}(2\nu_{0}+1)}\big\{J_{\nu_{0}+1}(\rho_{s})
 \left[N_{\nu_{0}+1}(\rho_{s})-N_{\nu_{0}-1}(\rho_{s})\right]
 -N_{\nu_{0}}(\rho_{s})\left[J_{\nu_{0}+2}(\rho_{s})
 -J_{\nu_{0}}(\rho_{s})\right]\big\} \,.
\label{Wgf}
\end{multline}
The fact that the integral on the right-hand side of
Eq.~(\ref{eqn:causality_bound_1a}) is positive semidefinite sets an upper
bound on the short-range parameter $K_{L,2}$.  We find that
\begin{equation}
 K_{L,2}\leq b_{L}(r)
\label{vdW_inequality}
\end{equation}
for any $r>R$.

\section{Impact on effective field theory}

\begin{figure}[ptb]
\begin{center}
\resizebox{100mm}{!}{\includegraphics{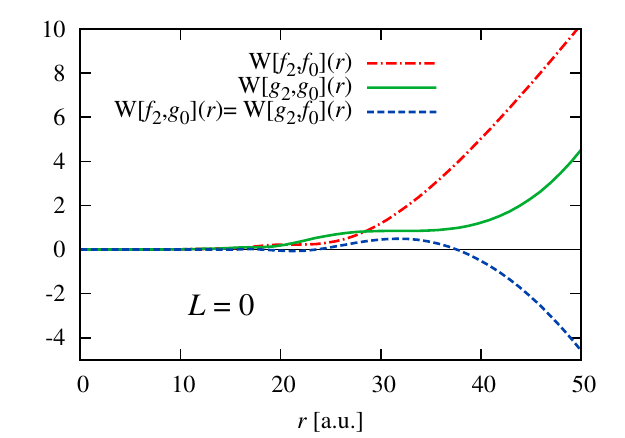}}
\end{center}
\caption{(Color online) Plot of $W[f_{0,2},f_{0,0}](r)$, $W[g_{0,2},g_{0,0}](r)$, and
$W[g_{0,2},f_{0,0}](r)$ as a function of $r$ for $L=0$ and $\beta_{6}=50$
(a.u.).}
\label{L0}
\end{figure}

\begin{figure}[ptb]
\begin{center}
\resizebox{100mm}{!}{\includegraphics{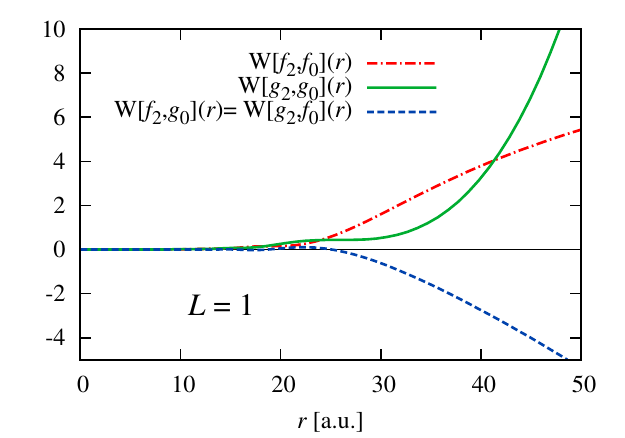}}
\end{center}
\caption{(Color online) Plot of $W[f_{1,2},f_{1,0}](r)$, $W[g_{1,2},g_{1,0}](r)$, and
$W[g_{1,2},f_{1,0}](r)$ as a function of $r$ for $L=1$ and $\beta_{6}=50$
(a.u.).}
\label{L1}
\end{figure}

\begin{figure}[ptb]
\begin{center}
\resizebox{100mm}{!}{\includegraphics{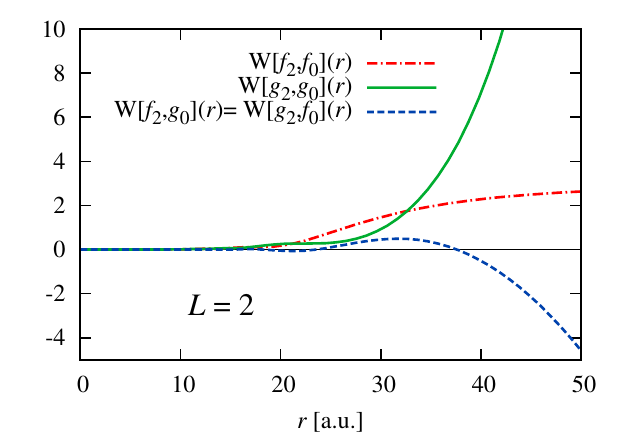}}
\end{center}
\caption{(Color online) Plot of $W[f_{2,2},f_{2,0}](r)$, $W[g_{2,2},g_{2,0}](r)$, and
$W[g_{2,2},f_{2,0}](r)$ as a function of $r$ for $L=2$ and $\beta_{6}=50$
(a.u.).}
\label{L2}
\end{figure}

In this section we discuss the impact of our causality bounds for an
effective field theory with short-range interactions and an attractive
$1/r^{6}$ tail.  In Fig.~\ref{L0} we plot the $L=0$ Wronskians
$W[f_{0,2},f_{0,0}]$, $W[g_{0,2},g_{0,0}]$, and $W[g_{0,2},f_{0,0}]$ for
$\beta_{6}=50$ (a.u.).  Figures~\ref{L1} and~\ref{L2} show the analogous
plots for $L=1$ and $L=2$, respectively.



We note that all of the Wronskian functions in Figs.~\ref{L0}, \ref{L1}, and
\ref{L2} vanish in the limit $r\rightarrow0$.  This stands in clear contrast
to what one finds for purely finite-range
interactions~\cite{Hammer:2009zh,Hammer:2010fw}.  In that case the effective
range parameter, $r_{L}$, satisfies the upper bound
\begin{equation}
 r_{L}\leq b_{L}^{\text{free}}(r) \,,
\label{rL_bound}
\end{equation}
where the function $b_{L}^{\text{free}}(r)$ is [\cf~Eq.~(60) in
Ref.~\cite{Hammer:2010fw}]
\begin{align}
 b_{L}^{\text{free}}(r) &= -\frac{2\Gamma(L-\frac{1}{2})\Gamma(L+\frac{1}{2})}
 {\pi}\left(\frac{r}{2}\right)^{-2L+1} \nonumber \\
 &-\frac{4}{L+\frac{1}{2}}\frac{1}{a_{L}}
 \left(\frac{r}{2}\right)^{2} \nonumber \\
 &+\frac{2\pi}{\Gamma(L+\frac{3}{2})\Gamma(L+\frac{5}{2})}
 \frac{1}{a_{L}^{2}}\left(\frac{r}{2}\right)^{2L+3} \,,
\label{2Ld_odd}
\end{align}
and $a_{L}$ is the scattering length.  Near $r=0$ the behavior of
$b_{L}^{\text{free}}(r)$ is
\begin{equation}
 b_{L}^{\text{free}}(r)=-\frac{2\Gamma(L-\frac{1}{2})\Gamma(L+\frac{1}{2})}{\pi}
 \left(\frac{r}{2}\right)^{-2L+1}+{O}(r^{2}) \,.
\end{equation}
We see that $b_{L}^{\text{free}}(r)$ diverges to negative infinity as
$r\rightarrow0$ for $L\geq1$.  The causality bound on $r_{L}$ also drives
$r_{L}$ to negative infinity for $L\geq1,$
\begin{equation}
 r_{L}\leq-\frac{2\Gamma(L-\frac{1}{2})\Gamma(L+\frac{1}{2})}{\pi}
 \left(\frac{r}{2}\right)^{-2L+1}+{O}(r^{2}) \,.
\end{equation}

For an effective field theory with local contact interactions, the range of
the interactions are controlled by the momentum cutoff scale.  No matter the
values for $a_{L}$ and $r_{L}$, it is not possible to take the momentum cutoff
scale arbitrarily high without violating the causality bound for channels with
angular momentum $L\geq1$.  For finite-range interactions with an additional
attractive or repulsive Coulomb tail, one finds the same leading
behavior~\cite{Koenig:2012bv}\footnote{To get this analogy, we use here
the normalization of the Coulomb-modified effective range expansion found in
Eq.~(28) of Ref.~\cite{Koenig:2012bv} and insert it in Eqs.~(64), (A.6), and
(A.7) of the same paper, which give the explicit expressions for the
Coulomb-modified causality bound functions for $L=0,1,2$.  The statement for
arbitrary $L$ then follows by generalization.}
\begin{equation}
 b_{L}^{\text{Coulomb}}(r)=-\frac{2\Gamma(L-\frac{1}{2})\Gamma(L+\frac{1}{2})}
 {\pi} \left(\frac{r}{2}\right)^{-2L+1}+{O}(r^{-2L}) \,,
\end{equation}
\ie, the only difference in the causality bound relation for the Coulomb-modified effective range is the subleading ${O}(r^{-2L})$ pole
term which is absent in the purely finite-range case.  Hence, also for an
effective field theory with contact interactions and long-range Coulomb tail, it
is not possible to take the momentum cutoff scale arbitrarily high for $L\geq1$
without violating the causality bound.

There is no such divergence in $b_{L}(r)$ at $r=0$ for the attractive
$1/r^{6}$ interaction.  For an effective field theory with contact
interactions and van der Waals tail, the causality bound does not impose
convergence problems as long as $K_{L,2}$ is less than or equal to zero.  This
holds true for any $L$.  There is no constraint from causality and unitarity
preventing one from taking the cutoff momentum to be arbitrarily large.  The key
difference between the van der Waals interaction and the Coulomb interaction is
that, when extended all the way to the origin, the attractive $1/r^{6}$
interaction is singular and the spectrum is unbounded below.  An essential
singularity appears at $r=0$, and both van der Waals wave functions $F_{L}$ and
$G_{L}$ vanish at the origin.

These exact same features appear in any attractive $1/r^{\alpha}$ interaction
for $\alpha>2$ in any number of spatial dimensions.  The same can be said
about an attractive $1/r^{2}$ interaction when the coupling constant is strong
enough to form bound states.  The key point is that in the zero-range
limit of these attractive singular potentials, the spectrum of bound states
extends to arbitrarily large negative energies.  As a consequence, the scattering
wave functions above threshold must vanish at the origin in order to satisfy
orthogonality with respect to all such bound-state wave functions localized near
the origin.  In all of these cases the function $b_{L}(r)$
remains finite as $r\rightarrow0$ for any $L$.  We conclude that for an
effective field theory with contact interactions and attractive singular power-law interactions, we can take the cutoff momentum arbitrarily large for any
$L$ without producing a divergence in the coefficient $K_{L,2}$ of the
short-range $K$ matrix.

\section{Quantum defect theory and the modified effective range expansion}

Up to now we have been discussing the short-range phase shift of $K$ matrix
for scattering relative to the van der Waals wave functions $F_{L}$ and $G_{L}%
$.  For power-law interactions $1/r^{\alpha}$ with $\alpha>2$, we also have
the option to define phase shifts relative to the Bessel functions of the free
wave equation.  The problem though is that the usual effective range
expansion,%
\begin{equation}
 k^{2L+1}\cot\delta_{L,d}(p)=-\frac{1}{a_{L}}+\frac{1}{2}r_{L}k^{2}
 +\sum_{n=0}^{\infty}(-1)^{n+1}\,\mathcal{P}_{L}^{(n)}k^{2n+4} \,,
\label{effectiverange}
\end{equation}
is spoiled by nonanalytic terms as a function of $k^{2}$.  For the van der
Waals interactions in the $L=0$ channel, the leading nonanalytic term is
proportional to $k^{3}$, and so the scattering parameters $a_{0}$ and $r_{0}$
are well defined, but the shape parameters $\mathcal{P}_{L}^{(n)}$ are not.
For $L=1$ the leading nonanalytic term is proportional to $k$ and so only
the scattering length $a_{1}$ is well defined.  For $L\geq2$ none of the
low-energy scattering parameters are well defined.  To resolve these
problems, a modified form of the effective range expansion is used which is
known as quantum-defect theory~\cite{RepProgPhys.46.167,PhysRevA.19.1485,PhysRevA.26.2441}.

In quantum-defect theory for attractive $1/r^{6}$ potentials, one defines an
offset for the phase shift~\cite{PhysRevA.80.012702},
\begin{equation}
 \eta_{L}=\frac{\pi}{2}(\nu-\nu_{0}) \,.
\end{equation}
The modified effective range expansion is then
\begin{equation}
 k^{2L+1}\cot\left({\delta}_{L}+2\eta_{L}\right) = -\frac{1}{A_{L}}
 +\frac{1}{2}R_{L}\,k^{2}+{O}\left(k^{4}\ln k\right) \,,
\label{eqn:generalized_ER_expansion_1a}
\end{equation}
where $A_{L}$ and $R_{L}$ are the generalized scattering length and effective
range parameters.  These definitions coincide with the usual scattering
length $a_{L}$ for $L=0,1$ and the usual effective range $r_{L}$ for $L=0$.
The generalized scattering length and effective range can be written in
terms of the short-range $K$ matrix parameters as
\begin{equation}
 A_{L}=\frac{\pi^{2}\beta_{6}^{2L+1}}{2^{4L+1}[\Gamma(\frac{L}2+\frac14)\Gamma
 (L+\frac32)]^{2}}\left[(-1)^{L}-\frac{1}{K_{L,0}}\right]
\label{eqn:generalized_scattering_length_1a}
\end{equation}
and
\begin{equation}
 R_{L}=-\frac{2^{4L+2}\Gamma\left(\frac{L}{2}+\frac{1}{4}\right)^{2}
 \Gamma\left(L+\frac{3}{2}\right)^{2}\beta_{6}^{-2L-1}}{\pi^{2}
 \left(K_{L,0}(-1)^{L}-1\right)^{2}}
 \left[\frac{\beta_{6}^{2}\left(K_{L,0}^{2}+1\right)}
 {4L^{2}+4L-3}-K_{L,2}\right] \,.
\label{eqn:generalized_effective_range_1a}
\end{equation}
From these results we see that the short-range parameter $K_{L,2}$ appears in
combination with $\beta_{6}^{2}$.  But in nearly all single-channel
scatterings between pairs of alkali-metal atoms, from the following equation,
\begin{align}
 K_{L,2} =
 &
 \frac{\beta_{6}^{2}}
 {4L^{2}+4L-3}
 \left\{
 1
 +\left[(-1)^{L}- A_{L}\frac{2^{4L+1}
  \Gamma\left(\frac{L}{2}+\frac{1}{4}\right)^{2}
  \Gamma\left(L+\frac{3}{2}\right)^{2}}{\pi^{2} \beta_{6}^{2L+1}} \right]^{-2}
 \right\}
\nonumber\\
 + & R_{L} A_{L}^{2}
  \frac{2^{4L}\Gamma\left(\frac{L}{2}+\frac{1}{4}\right)^{2}
   \Gamma\left(L+\frac{3}{2}\right)^{2}}{\pi^{2}\beta_{6}^{2L+1}
 }
 \left[(-1)^{L}- A_{L}\frac{2^{4L+1}
  \Gamma\left(\frac{L}{2}+\frac{1}{4}\right)^{2}
  \Gamma\left(L+\frac{3}{2}\right)^{2}}{\pi^{2} \beta_{6}^{2L+1}} \right]^{-2},
\label{eq:K_2_effective_range}
\end{align}
one quantitatively finds that $K_{L,2}$ is at
least one order of magnitude smaller than $\beta_{6}^{2}$.  This separation
of scales is the reason for the approximate universality found in
Refs.~\cite{CalleCordon:2010a,RuizArriola:2011a}.

The dominance of $\beta_6^{2n}$ over the subleading coefficients $K_{L,2n}$
in Eq.~(\ref{eq:K_L_expansion}) for $n \geq 1$ holds for nearly all cases of
single-channel scattering between alkali-metal atoms
\cite{PhysRevA.58.4222,PhysRevA.64.010701}.
This phenomenological fact explains the absence of short-distance length scales
in the universality relation in Eq.~(\ref{arriola}).
Furthermore, Gao has shown that when short-range interactions arise from a
repulsive central potential, the fact that the $K$ matrix is nearly
independent of energy means that the $K$ matrix is also nearly independent of
angular momentum $L$ \cite{PhysRevA.64.010701}.  This produces a surprisingly
rich class of universal physics for single-channel van der Waals interactions
where $K_{L,2n}$ is negligible compared to $\beta_6^{2n}$ for all $L$,
and $K_{L,0}$ is approximately the same for all $L$.  Therefore $\beta_6$
and the $s$-wave scattering length will determine, to a good approximation,
the threshold scattering behavior for all values of $L$.

\section{Causal range for single-channel scattering}

We have shown that for negative $K_{L,2}\leq0$, the range $R$ of the
short-range interaction can be taken all the way down to zero.  But when
$K_{L,2}$ is positive, there is a constraint on $R$ and we use
Eq.~(\ref{vdW_inequality}) to determine a minimum value for $R$.  We call
this minimum range the causal range $R^{b}$, and we determine $R^{b}$ as the
solution to the equation%
\begin{equation}
 K_{L,2}=b_{L}(R^{b}) \,.
\label{causal_constraint}
\end{equation}
As pointed out in Ref.~\cite{Koenig:2012bv}, one can show \textit{a priori} that
$b_{L}(r)$ is a monotonically increasing function of $r$.  Therefore, if a
real solution to Eq.~(\ref{causal_constraint}) exists, then it is unique.  If,
however, there is no real solution, then there is no constraint on the
interaction range and we define $R^{b}$ to be zero.  For an effective field
theory with contact interactions and van der Waals tail, the cutoff momentum
can be made as large as $\sim1/R^{b}$ before the causality bound is violated.

In the following analysis we extract the single-channel $s$-wave effective range
parameters $a_{0}$ and $r_{0}$ for several different pairs of alkali-metal atoms
$^{7}$Li, $^{23}$Na, and $^{133}$Cs in singlet and triplet channels.  The
data is taken from Refs.~\cite{PhysRevA.50.4827,
PhysRevA.53.234,PhysRevA.50.3177,PhysRevA.59.1998,PhysRevA.58.4222}.  The
reduced masses for $^{7}$Li$_{2}$, $^{23}$Na$_{2}$, and $^{133}$Cs$_{2}$ are
$\mu=6394.7,20954,121100$ (a.u.), respectively.  The van der Waals coupling
constants for $^{7}$Li$_{2}$, $^{23}$Na$_{2}$, and $^{133}$Cs$_{2}$ are
$C_{6}=1388,1472,7020$ (a.u.).  We calculate the corresponding $K$-matrix
parameters using Eq.~\eqref{eqn:generalized_scattering_length_1a} and
Eq.~\eqref{eq:K_2_effective_range} and then compute the resulting
causal ranges.  We recall that for $L=0$ we simply have $A_{0}=a_{0}$ and
$R_{0}=r_{0}$.  The results for the scattering parameters and causal ranges
are given in columns \Rmnum{2}, \Rmnum{5} and \Rmnum{6} of
Table~\ref{causalranges}.  The discrepancies in $R_{0}$ are due to the fact that
in the analytic studies in Refs.~\cite{PhysRevA.59.1998,PhysRevA.58.4222}
$K_{0,2}$ is neglected, while the numerical calculations of
Refs.~\cite{PhysRevA.50.4827,PhysRevA.53.234,PhysRevA.50.3177} include the
short-range contribution from $K_{0,2}$.

In column V of Table~\ref{causalranges}, we present an approximate range
for $K_{0,2}$ for each atomic pair using the values for $R_0$ in columns III
and IV.  Since $K_{0,2}$ is positive, we cannot go all the way to the
zero-range limit.  However, in each case $K_{0,2}$ is at least one order of
magnitude smaller than $\beta_{6}^{2}$.\footnote{Note that $K_{L,2}$ has the
dimension of an area (in the appropriate atomic units).}  Although we cannot
take the zero-range limit, the causal ranges are small in comparison to
$\beta_{6}$. In each case $R^{b}$ is less than one-third the size of
$\beta_{6}$.  Hence one can probe these interactions in a van der Waals
effective field theory with cutoff momentum up to roughly three times
$1/\beta_{6}$ without violating the causality bound.

\begin{table}[ptb]
\caption{Scattering parameters and causal ranges for $s$-wave scattering of
$^{7}$Li$,^{23}$Na$,$ and $^{133}$Cs pairs.  The scattering data collection is
taken from Ref.~\cite{PhysRevA.59.1998}. In columns \Rmnum{1} and \Rmnum{4}
the scattering data for $^{7}$Li are from Ref.~\cite{PhysRevA.53.234},
the scattering data for $^{23}$Na are from
Refs.~\cite{PhysRevA.50.4827,PhysRevA.53.234}, and data for $^{133}$Cs are from
Ref.~\cite{PhysRevA.50.3177}. In column \Rmnum{3}  the effective range
parameters, $R_{0}$, are calculated analytically in
Refs.~\cite{PhysRevA.59.1998, PhysRevA.58.4222}. In column \Rmnum{4}, the
$R_{0}$ are obtained from numerical calculations.  The scattering parameters in
columns \Rmnum{2} and \Rmnum{5} are calculated using
Eqs.~(\ref{eqn:generalized_scattering_length_1a})
and~(\ref{eq:K_2_effective_range}), and the causal ranges in column \Rmnum{6}
are obtained from Eq.~(\ref{causal_constraint}).}
\label{causalranges}
{\footnotesize{
\begin{tabular}
[c]{c|c|c|c|c|c|c|c|c}\hline\hline
\multicolumn{3}{c|}{} & \Rmnum{1} & \Rmnum{2}  & \Rmnum{3}  & \Rmnum{4}
& \Rmnum{5}  & \Rmnum{6}\\
\hline\hline
Atoms & State & $\beta_{6}$ & $A_{0}$ & $K_{0,0}$ & $R_{0}$ & $R_{0}$ &
 $K_{0,2}$ & $R^b$
\\
\hline
$^{7}$Li--$^{7}$Li & $^{1}\Sigma_{g}$ & 64.9097 & 36.9 & -5.282 & 66.3 & 66.5 &
 2 $\sim$ 124 & 7 $\sim$ 19
\\
$^{7}$Li--$^{7}$Li & $^{3}\Sigma_{u}$ & 64.9097 & -17.2 & 0.643 & 1006.3
 & 1014.8 & 0 $\sim$ 17 & 3 $\sim$ 25
\\
\hline
$^{23}$Na--$^{23}$Na & $^{1}\Sigma_{g}$ & 88.624 & 34.936 & 5.705 & 187.317
& 187.5 & 0 $\sim$ 86 & 4 $\sim$ 20
\\
$^{23}$Na--$^{23}$Na & $^{3}\Sigma_{u}$ & 88.624 & 77.286 & -1.213
& 62.3756 & 62.5 &  2 $\sim$ 13 & 16 $\sim$ 24
\\
\hline
$^{133}$Cs--$^{133}$Cs & $^{1}\Sigma_{g}$ & 203.62 & 68.216 & 3.365 & 624.013
& 624.55 & 0 $\sim$ 146 & 7 $\sim$ 45\\
\hline\hline
\end{tabular}}}
\end{table}

\section{Causal range near a magnetic Feshbach resonance}

In Ref.~\cite{Gao:2011a} the multichannel problem of scattering around a
magnetic Feshbach resonance is reduced to a description by an effective
single-channel $K$ matrix that depends on the applied magnetic field $B$.
\ The behavior around the resonance is described by several parameters.
$B_{0,L}$ is the position of the resonance, while $g_{\text{res}}$
parametrizes the width of the Feshbach resonance. $\ K_{L}^{\text{bg}}$ is a
background value for the $K$ matrix, and the scale $d_{B,L}$ is introduced to
define a dimensionless magnetic field.  We write the effective single-channel
$K$ matrix as
\begin{equation}
 \hat{K}_{L}^{\text{eff}}(k,B)=-K_{L}^{\text{bg}}
 \left[1+\frac{g_{\text{res}}}{k^{2}\beta_{6}^{2}-g_{\text{res}}
 \left(B_{s}+1\right)}\right] \,,
\label{kL}%
\end{equation}
with
\begin{equation}
 B_{s}=\frac{\left(B-B_{0,L}\right)}{d_{B,L}} \,.
\label{Bs}
\end{equation}
The parametrization given above corresponds to Eq. (18) in
Ref.~\cite{Gao:2011a}.  Note that we have changed the notation slightly and
are using a different sign convention.

By expanding the right-hand side of Eq.~(\ref{kL}) in $k^{2}$, it is
straightforward to determine the $K$ matrix expansion parameters $K_{L,0}$ and
$K_{L,2}$.  A short calculation yields that
\begin{equation}
 K_{L,0}^{\text{eff}}=-K_{L}^{\text{bg}}\left(1+\frac{1}{B_{s}+1}\right) \,,
\end{equation}
\begin{equation}
 K_{L,2}^{\text{eff}}=\frac{\beta_{6}^{2}K_{L}^{\text{bg}}}
 {g_{\text{res}}(B_{s}+1)^{2}} \,.
\label{KL2}
\end{equation}
As noted in Ref.~\cite{Gao:2011a}, the parameters $K_{L}^{\text{bg}}$ and
$g_{\text{res}}$ are constrained by the condition
\begin{equation}
 K_{L}^{\text{bg}}g_{\text{res}}<0 \,.
\end{equation}
From this we directly see that $K_{L,2}^{\text{eff}}$ given by Eq.~(\ref{KL2})
is always negative.  From the causality bound in Eq.~(\ref{vdW_inequality})
it follows that where this effective single-channel description is applicable
and correctly captures the entire energy dependence of the short-range
$K$ matrix, the causal range will be zero when the interaction is tuned close
to a Feshbach resonance.

\section{Summary and discussion}

In this paper we have analyzed two-body scattering with arbitrary short-range
interactions plus an attractive $1/r^{6}$ tail.  We derived the constraints
of causality and unitarity for the short-range $K$ matrix,
\begin{equation}
 \hat{K}_{L}=\tan\delta_{L}^{(\text{short})}(k)
 =\sum_{n=0}^{\infty}K_{L,2n}\,k^{2n} \,.
\end{equation}
For any $r$ larger than the range of the short-range interactions, $R$, we
find that $K_{L,2}$ satisfies the upper bound
\begin{equation}
 K_{L,2}\leq b_{L}(r) \,,
\end{equation}
where $b_{L}(r)$ is
\begin{align}
 b_{L}(r) &=\frac{\pi}{4}W[f_{L,2},f_{L,0}](r)+\frac{\pi}{4}K_{L,0}^{2}
 W[g_{L,2},g_{L,0}](r) \nonumber \\
 &-\frac{\pi}{4}K_{L,0}\left\{\vphantom{\frac{1}{2}}W[g_{L,2},f_{L,0}](r)
 +W[f_{L,2},g_{L,0}](r)\right\} \,,
\end{align}
and the Wronksians are given in Eq.~(\ref{Wff}), (\ref{Wgg}), and (\ref{Wgf}).

In clear contrast with the case for finite-range interactions
only~\cite{Hammer:2009zh,Hammer:2010fw} or with Coulomb
tails~\cite{Koenig:2012bv}, the function $b_{L}(r)$ does not diverge but rather
vanishes as $r\rightarrow 0$ for all $L$.  When $K_{L,2}\leq0$, there is no
constraint derived from causality and unitarity that prevents the use of an
effective field theory with zero-range contact interactions plus an attractive
$1/r^{6}$ tail.  This holds true for any angular momentum value $L$.  For the
phenomenologically important case of a multichannel system near a
magnetic Feshbach resonance, the effective value for $K_{L,2}$ is
negative and so the short-range
interaction can be taken to have zero range.

The van der Waals interaction is qualitatively different from the Coulomb
interaction where $b_{L}^{\text{Coulomb}}(r)$ diverges for $L\geq1$.  The key difference is
that both van der Waals wave functions $F_{L}$ and $G_{L}$ vanish at the origin.
This phenomenon also occurs for an attractive $1/r^{\alpha}$ interaction for
$\alpha>2$ in any number of spatial dimensions.  It is also valid for an
attractive $1/r^{2}$ interaction when the coupling constant is strong enough
to form bound states.  For an effective field theory with contact
interactions and attractive singular power-law tail, the cutoff momentum can
be made arbitrarily large for any $L$ without producing a divergence in the
coefficient $K_{L,2}$ of the short-range $K$ matrix.

When $K_{L,2}$ is positive, there is a lower bound on the range of the
short-range interactions.  We define the causal range $R^{b}$ as this minimum
value for the range, given by the condition%
\begin{equation}
 K_{L,2}=b_{L}(R^{b}) \,.
\end{equation}
We have analyzed several examples of $s$-wave scattering in alkali-metal atoms in
Table~\ref{causalranges}.  We find that the $K_{L,2}$ is at least one order
of magnitude smaller than $\beta_{6}^{2}$.  As a result we find that the
causal ranges are small in comparison with $\beta_{6}$.

In summary, we find that $\beta_6$ dominates over distance scales parametrizing
the short-range interactions.  The origin of this van der Waals universality
can be explained by two facts.  The first fact is the phenomenological
observation that, in single-channel scattering between alkali-metal atoms, there
is a significant separation
between the typical length scales of the short-distance physics and $\beta_6$.
This can be seen by the small size of the short-range parameter $K_{L,2}$
compared with $\beta_{6}^{2}$.  As Gao has shown, this also leads to
the approximate universal relation that $K_{L,0}$ is the same for all $L$
\cite{PhysRevA.64.010701}.  Therefore, to a good approximation,
$\beta_6$ and the $s$-wave scattering length will determine the threshold
scattering behavior for all values of $L$.  For the multichannel case
near a magnetic Feshbach resonance, we find that the
effective $K_{L,2}^{\text{eff}}$ is no
longer negligible.  However, $K_{L,2}^{\text{eff}}$ is negative, and this means
that there is no constraint from causality preventing the zero-range limit for
the short-distance interactions.

The second fact underlying the van der Waals universality is that
the zero-range limit of short-distance interactions is
well behaved with regard to scattering near threshold.  We note, however, that
there is still no scale-invariant limit for $L \geq 1$ since the effective range
parameter will diverge to negative infinity as $\beta_6$ goes to zero.  This can
be seen from the $\beta_6^{-2L+1}$ behavior with negative coefficient for $L \geq 1$
in Eq.~(\ref{eqn:generalized_effective_range_1a}).

The analysis in this paper should be useful in developing an effective field
theory with an attractive $1/r^{6}$ tail and contact interactions.  Similarly,
one can also construct effective field theories for other attractive singular
potentials $1/r^{\alpha}$ for $\alpha\geq2$.  These effective field theories
could be used to investigate the Efimov effect and other low-energy phenomena in
a model-independent way.

\begin{acknowledgments}
This research was supported in part by the US Department of Energy under
Contract No. DE-FG02-03ER41260, by the BMBF under Contract No. 05P12PDFTE,
and by DFG and NSFC (CRC 110).  S.E. was supported by a Turkish Government
Ministry National Education Fellowship, and S.K. was supported by the
\textquotedblleft Studien\-stiftung des deutschen Volkes\textquotedblright
and by the Bonn-Cologne Graduate School of Physics and Astronomy.  We also
thank the Institute for Nuclear Theory for hosting the authors during the \textquotedblleft Light nuclei from first principles\textquotedblright program.
\end{acknowledgments}

\appendix

\section{Van der Waals wave functions}

\label{append:analytic_solution_vdWaals_eq}

In this section we derive the van der Waals wave functions $F_{L}$ and $G_{L}$,
following the steps in Ref.~\cite{PhysRevA.58.1728}.  We first redefine the
radial function as $U_{L}(r)=\sqrt{r_{s}}Z(\rho_{s})$. This rearrangement
puts Eq.~(\ref{pure_vdW}) into the form of an inhomogeneous Bessel equation,
\begin{equation}
 \mathcal{L}_{\nu_{0}}Z(\rho_{s})=\left[\rho_{s}^{2}\frac{d^{2}}{d\rho_{s}^{2}}
 +\rho_{s}\frac{d}{d\rho_{s}}-\nu_{0}^{2}+\rho_{s}^{2}\right]
 Z(\rho_{s})=-\frac{k_{s}^{2}}{8}\frac{Z(\rho_{s})}{\rho_{s}} \,,
\label{eqn:inhomogeneous_Bessel_equ_1a}
\end{equation}
with
\begin{equation*}
\nu_{0}=\frac{1}{4}(2L+1).
\end{equation*}
The idea, introduced in Ref.~\cite{PhysRevA.50.2841}, is now to consider
$Z_{\nu}(\rho_{s})$ as a series expansion of solutions,
\begin{equation}
 Z(\rho_{s})=\sum_{n=0}^{\infty}k_{s}^{2n}\,\varphi^{(n)}(\rho_{s}) \,,
\label{eqn:Z_rho_expansion_1a}
\end{equation}
and to use perturbation theory to obtain a solution for $Z_{\nu}(\rho_{s})$.
Substituting Eq.~(\ref{eqn:Z_rho_expansion_1a}) into
Eq.~(\ref{eqn:inhomogeneous_Bessel_equ_1a}) leads to an infinite number of
differential equations,
\begin{align}
 \mathcal{L}_{\nu_{0}}\,\varphi^{(0)}(\rho_{s})\,+k_{s}^{2}\,
 &\left[\mathcal{L}_{\nu_{0}}\,\varphi^{(1)}(\rho_{s})\,+\frac{1}{8\rho_{s}}
 \,\varphi^{(0)}(\rho_{s})\right] \nonumber \\
 &+k_{s}^{4}\,\left[\mathcal{L}_{\nu_{0}}\,\varphi^{(2)}(\rho_{s})
 \,+\frac{1}{8\rho_{s}}\,\varphi^{(1)}(\rho_{s})\right] +\cdots = 0 \,.
\label{eqn:inhomogeneous_Bessel_equ_2a}
\end{align}
The zeroth-order differential equation is homogenous, while all other orders
are inhomogeneous. This procedure generates a secular perturbation in all
inhomogeneous differential equations as well as driving terms.  The secular
terms here refer to the solutions of the zeroth-order differential equation,
which are Bessel functions.

Following Ref.~\cite{PhysRevA.58.1728}, we introduce a function
$Z_{\nu}(\rho_{s})$ which has an expansion in terms of Bessel functions with
momentum-dependent coefficients,
\begin{equation}
 Z_{\nu}(\rho_{s})=\sum_{m=-\infty}^{\infty}b_{m}(k_{s})
 \,\mathcal{J}_{\nu+m}(\rho_{s}) \,.
\label{Zv}
\end{equation}
We insert this as an ansatz into Eq.~(\ref{eqn:inhomogeneous_Bessel_equ_1a})
with $\nu$ yet to be determined.  Here $\mathcal{J}_{n}$ denotes collectively
the Bessel functions of the first and second kind, $J_{n}$ and $N_{n}$.
Substitution of Eq.~(\ref{Zv}) into Eq.~(\ref{eqn:inhomogeneous_Bessel_equ_1a})
yields a three-term recurrence relation for the $b_{m}$ functions with
$-\infty<m<\infty$,
\begin{equation}
 \lbrack(\nu+m)^{2}-\nu_{0}^{2}]b_{m}(k_{s})+\frac{k_{s}^{2}}{16(m+\nu-1)}
 b_{m-1}(k_{s})+\frac{k_{s}^{2}}{16(m+\nu+1)}b_{m+1}(k_{s})=0 \,.
\label{recurrence}
\end{equation}
Solving these equations for $b_{m}(k_{s})$ yields
\begin{equation}
 b_{m}(k_{s})=\left(-1\right)^{m}\left(\frac{k_{s}}{4}\right)^{2m}
 \frac{\Gamma(\nu)\Gamma(\nu-\nu_{0}+1)\Gamma(\nu+\nu_{0}+1)}
 {\Gamma(\nu+m)\Gamma(\nu-\nu_{0}+m+1)\Gamma(\nu+\nu_{0}+m+1)}\,c_{m}(\nu)
\label{bm}
\end{equation}
and
\begin{equation}
 b_{-m}(k_{s})=\left(-1\right)^{m}\left(\frac{k_{s}}{4}\right)^{2m}
 \frac{\Gamma(\nu-m+1)\Gamma(\nu-\nu_{0}-m)\Gamma(\nu+\nu_{0}-m)}
 {\Gamma(\nu+1)\Gamma(\nu-\nu_{0})\Gamma(\nu+\nu_{0})}\,c_{m}(-\nu)
\label{bmm}
\end{equation}
for $m\geq0$.  The functions $c_{m}(\pm\nu)$ are defined as
\begin{equation}
 c_{m}(\pm\nu)=\prod_{s=0}^{m-1}Q(\pm\nu+s)\,b_{0}(k_{s}) \,,
\end{equation}
where $Q(\nu)$ is given by
\begin{equation}
 Q(\nu)=\cfrac{1}{1-\cfrac{k_{s}^{2} }{16(\nu+1)[(\nu+1)^2-\nu_o^2]
 (\nu+2)[(\nu+2)^2-\nu_o^2]}Q(\nu+1)} \,.
\label{Q}
\end{equation}
The coefficient $b_{0}(k_{s})$ only determines the overall normalization and
is simply set to one in the following.  Equation~(\ref{recurrence}) for $m=0$
determines the shift $\nu$ in the order of the Bessel functions.  We
determine $\nu$ using the constraint
\begin{equation}
 (\nu^{2}-\nu_{0}^{2})-\frac{Q(-\nu)}{16^{2}\nu(\nu-1)[(\nu-1)^{2}-\nu_{0}^{2}]}
 \ k_{s}^{4}-\frac{Q(\nu)}{16^{2}\nu(\nu+1)
 [(\nu+1)^{2}-\nu_{0}^{2}]}\ k_{s}^{4}=0 \,.
\label{v}
\end{equation}
In general there are several roots which become complex beyond a critical
scaled momentum $k_{s}$, and one must be careful to choose the physical
solution.  For a detailed discussion of this point, see
Refs.~\cite{PhysRevA.58.1728,PhysRevA.80.012702}.

Choosing either $\mathcal{J}_{n}=J_{n}$ or $\mathcal{J}_{n}=N_{n}$ already
yields a pair of linearly independent solutions. However, in order to get a
pair with energy-independent normalization as $r_{s}\rightarrow0$ (which
ensures analyticity in the energy), we furthermore define
\begin{equation}
 X_{L}(k_{s})=\cos\eta_{L}\sum_{m=-\infty}^{\infty}(-1)^{m}b_{2m}(k_{s})
 -\sin\eta_{L}\sum_{m=-\infty}^{\infty}(-1)^{m}b_{2m+1}(k_{s})
\label{X_L}
\end{equation}
and
\begin{equation}
 Y_{L}(k_{s})=\sin\eta_{L}\sum_{m=-\infty}^{\infty}(-1)^{m}b_{2m}(k_{s})
 +\cos\eta_{L}\sum_{m=-\infty}^{\infty}(-1)^{m}b_{2m+1}(k_{s}) \,,
\label{Y_L}
\end{equation}
with
\begin{equation*}
\eta_{L}=\frac{\pi}{2}(\nu-\nu_{0}) \,.
\end{equation*}
Combining everything, we arrive at the van der Waals wave functions,
\begin{equation}
 {F}_{L}(k,r)=\frac{r_{s}^{1/2}}{X_{L}^{2}(k_{s})+Y_{L}^{2}(k_{s})}
 \left[X_{L}(k_{s})\sum_{m=-\infty}^{\infty}b_{m}(k_{s})
 \,J_{\nu+m}\left(\rho_{s}\right)
 -Y_{L}(k_{s})\sum_{m=-\infty}^{\infty}b_{m}(k_{s})
 \,N_{\nu+m}\left(\rho_{s}\right)\right] \,,
\end{equation}
\begin{equation}
 {G}_{L}(k,r)=\frac{r_{s}^{1/2}}{X_{L}^{2}(k_{s})+Y_{L}^{2}(k_{s})}
 \left[X_{L}(k_{s})\sum_{m=-\infty}^{\infty}b_{m}(k_{s})
 \,N_{\nu+m}\left(\rho_{s}\right)+Y_{L}(k)\sum_{m=-\infty}^{\infty}b_{m}(k_{s})
 \,J_{\nu+m}\left(\rho_{s}\right)\right] \,.
\end{equation}

\section{Low-energy expansions}
\label{low_energy_expansion}

In this appendix we expand all functions relating to the van der Waals
wave functions in powers of momentum.  We first consider $\nu$, the shift in
the order of the Bessel functions in Eq.~(\ref{F_L}) and Eq.~(\ref{G_L}).
Using Eq.~$(\ref{v})$ in Appendix~\ref{append:analytic_solution_vdWaals_eq},
we find
\begin{equation}
 \nu=\nu_{0}-\frac{3}{2^{8}\nu_{0}(4\nu_{0}^{2}-1)(\nu_{0}^{2}-1)}k_{s}^{4}
 +{O}(k_{s}^{8}) \,,
\end{equation}
where $\nu_{0}=(2L+1)/4$.  Using the expansion in Eqs.~(\ref{bm}),
(\ref{bmm}), and~(\ref{Q}), we get
\begin{equation}
 b_{m}(k_{s})=\left(-1\right)^{m}\frac{\Gamma(\nu_{0})\Gamma(2\nu_{0}+1)}
 {m!\,\Gamma(\nu_{0}+m)\Gamma(2\nu_{0}+m+1)}\left(\frac{k_{s}}{4}\right)^{2m}
+{O}(k_{s}^{2m+2})
\end{equation}
and
\begin{equation}
 b_{-m}(k_{s})=\frac{\Gamma(\nu_{0}-m+1)\Gamma(2\nu_{0}-m)}{m!\,\Gamma(\nu_{0}
 +1)\Gamma(2\nu_{0})}\left(\frac{k_{s}}{4}\right)^{2m}+{O}(k_{s}^{2m+2})
\end{equation}
for $m\geq0$.  Substituting these expressions into Eqs.~(\ref{X_L})
and~(\ref{Y_L}) we obtain
\begin{equation}
 X_{L}(k_{s})=1+{O}(k_{s}^{4}) \,,
\end{equation}
\begin{equation}
 Y_{L}(k_{s})=-\left[  \frac{\Gamma(\nu_{0})\Gamma(2\nu_{0}-1)}{\Gamma(\nu_{0}
 +1)\Gamma(2\nu_{0})}+\frac{\Gamma(\nu_{0})\Gamma(2\nu_{0}+1)}
 {\Gamma(\nu_{0}+1)\Gamma(2\nu_{0}+2)}\right]\left(\frac{k_{s}}{4}\right)^{2}
 +{O}(k_{s}^{4}) \,.
\end{equation}

\end{document}